\documentclass[conference]{IEEEtran}
\IEEEoverridecommandlockouts
\usepackage{cite}
\usepackage{amsmath,amssymb,amsfonts}
\usepackage{algorithmic}
\usepackage{graphicx}
\usepackage{balance}
\usepackage{textcomp}
\usepackage{xcolor}
\usepackage{booktabs}
\usepackage{enumitem}
\setlist[itemize]{noitemsep, topsep=0pt}

\usepackage{lipsum}
\usepackage{multirow}
\def\BibTeX{{\rm B\kern-.05em{\sc i\kern-.025em b}\kern-.08em
    T\kern-.1667em\lower.7ex\hbox{E}\kern-.125emX}}

\usepackage[toc, acronym, nopostdot, nonumberlist]{glossaries}
    \newacronym{api}{API}{Application Programming Interface}
\newacronym{der}{DER}{distributed energy resource}
\newacronym{adms}{ADMS}{advanced distribution management system}
\newacronym{derms}{DERMS}{distributed energy resource management system}
\newacronym{dso}{DSO}{Distribution System Operator}
\newacronym{flsir}{FLSIR}{Fault Location, Isolation, and Service Restoration}
\newacronym{vvo}{VVO}{Volt-VAR Optimization}
\newacronym{dec}{DEC}{distributed energy coordinator}
\newacronym{fa-dec}{FA-DEC}{fairness-aware distributed energy coordinator}
\newacronym{pv}{PV}{photovoltaic}
\newacronym{sda}{SDA}{switch delimited area}
\newacronym{iot}{IoT}{Internet of Things}
\newacronym{ot}{OT}{Operational Technology}
\newacronym{pcc}{PCC}{point of common coupling}
\newacronym{mv}{MV}{medium voltage}
\newacronym{cim}{CIM}{Common Information Model}
\newacronym{json}{JSON}{JavaScript Object Notation}

\begin{document}

\newcommand\blfootnote[1]{%
\begingroup
\renewcommand\thefootnote{}\footnote{#1}%
\addtocounter{footnote}{-1}%
\endgroup
}
\renewcommand{\footnoterule}{%
\kern -3pt
\hrule width 0.29 \textwidth height 0.5pt
\kern 1pt
}

% \def\footnoterule{\kern-3\p@
%   \hrule \@width 2in \kern 2.6\p@} % the \hrule is .4pt high
% \makeatother

\title{Risk-Aware Planning of Power Distribution Systems Using Scalable Cloud Technologies
\vspace{-0.25 cm}
% \thanks{This material is based on work supported by the U.S. Department of Energy (DOE), Advanced Grid Research Program. Pacific Northwest National Laboratory is operated for DOE by the Battelle Memorial Institute under Contract DE-AC05-76RL01830.
% }
}
% \vspace{-0.75 cm}
\author{\IEEEauthorblockN{
Shiva Poudel\IEEEauthorrefmark{1},  
Poorva Sharma\IEEEauthorrefmark{1},
Abhineet Parchure\IEEEauthorrefmark{2},
Daniel Olsen\IEEEauthorrefmark{2},
Sayantan Bhowmik\IEEEauthorrefmark{2},
Tonya Martin\IEEEauthorrefmark{1}, \\
Dylan Locsin\IEEEauthorrefmark{2}, and
Andrew P. Reiman\IEEEauthorrefmark{1} 
 }
\IEEEauthorblockA{\IEEEauthorrefmark{1}Pacific Northwest National Laboratory (PNNL), Richland, WA, USA}
    \IEEEauthorblockA{\IEEEauthorrefmark{2}Amazon Web Services (AWS), Seattle, WA, USA}
\vspace{-0.5 cm}
% Email: shiva.poudel@pnnl.gov
}

\IEEEoverridecommandlockouts
\IEEEpubid{\makebox[\columnwidth]{978-1-5386-5541-2/18/\$31.00~\copyright2018 IEEE \hfill} \hspace{\columnsep}\makebox[\columnwidth]{ }}

\maketitle
\begin{abstract}
The uncertainty in distribution grid planning is driven by the unpredictable spatial and temporal patterns in adopting electric vehicles (EVs) and solar photovoltaic (PV) systems. This complexity, stemming from interactions among EVs, PV systems, customer behavior, and weather conditions, calls for a scalable framework to capture a full range of possible scenarios and analyze grid responses to factor in compound uncertainty. Although this process is challenging for many utilities today, the need to model numerous grid parameters as random variables and evaluate the impact on the system from many different perspectives will become increasingly essential to facilitate more strategic and well-informed planning investments. We present a scalable, stochastic-aware distribution system planning application that addresses these uncertainties by capturing spatial and temporal variability through a Markov model and conducting Monte Carlo simulations leveraging modular cloud-based architecture. The results demonstrate that 15,000 power flow scenarios generated from the Markov model are completed on the modified IEEE 123-bus test feeder, with each simulation representing an 8,760-hour time series run, all in under an hour. The grid impact extracted from this huge volume of simulated data provides insights into the spatial and temporal effects of adopted technology, highlighting that planning solely for average conditions is inadequate, while worst-case scenario planning may lead to prohibitive expenses.

% We develop a stochastic-aware distribution system planning application and demonstrate the application using scalable cloud services. By providing rich datasets and enabling detailed analysis of improbable outcomes, this framework supports advanced decision-making in distribution-grid planning and operations amidst the transition to a decentralized energy paradigm. 
% Our research serves as a proof-of-concept illustrating how cloud-empowered stochastic modeling enhances distribution-grid planning decisions, enabling operators to navigate a landscape characterized by multiple sources of uncertainty.
\end{abstract}

\section{Introduction}
The global imperative for transitioning to a net-zero future has spurred a significant consumer interest in embracing renewable energy solutions, notably solar photovoltaic (PV) systems and electric vehicles (EVs) \cite{sharda2024electric}. 
However, the location and capacity uncertainty of DER installations, the influence of weather patterns, and varying customer behavior introduce variability in power flow patterns, complicating the planning and management of distribution networks. 

Stochastic-optimization-based models are explored in the literature to account for the uncertainty during planning for EV charging station infrastructure and power distribution network \cite{wang2019expansion}, \cite{de2017joint}. However, these approaches use approximate models for power flow and don't capture detailed modeling of components such as service transformers.
\textcolor{black}{Advanced algorithms leveraging transactive markets \cite{10599533}, chance-constrained programming \cite{9316189}, and correlated probability distributions \cite{wanninayaka2024probabilistic} have been proposed for joint EV and PV planning; however, these approaches are primarily designed for short-term planning and are not directly applicable to extended planning horizons.}
% \cite{10599533, 9316189} - short term planning problem, not applicable for longer horizon...
Important sampling within the Monte-Carlo (MC) framework is studied in power systems planning studies to reduce the search space in stochastic optimization  \cite{urgun2019importance}. However, given the emphasis on the worst-case scenario, these methods may lead to extremely conservative and expensive planning solutions as the worst-case scenario usually occurs with a very small probability \cite{9942328}. Furthering the complexity, the choice of a simplified model can amplify the error produced from lower temporal resolution (or reduced scenario) \cite{parchure2016investigating, marcy2022comparison}.
Therefore, it is important to consider a full range of possible future scenarios to factor in compounded uncertainty. MC-based approaches are well suited for planning studies with uncertainty in decision variables \cite{8848458}. However, the major challenge of the MC technique is that to achieve a satisfactory probabilistic distribution approximation accuracy, many samples need to be generated, which is time, computation, and data-intensive---local and inelastic computing resources provisioned by power system organizations today are inadequate for such demands~\cite{zhang2022practical}.
Recently, advanced cloud-based solutions tailored to utility needs have emerged, providing faster grid analytics, delivering actionable insights from vast amounts of data, with the potential to offer scalable computing without the need for prohibitively expensive supercomputing infrastructure \cite{cloud, zhang2022practical}. This shift is further emphasized by a recent statement from the White House, which highlights the potential of advanced computing and software solutions to accelerate the integration of clean energy into the grid \cite{whitehouse}.

We propose a scalable and extensible framework that leverages GridAPPS-D and Amazon Web Services (AWS) to share grid models and data among applications and provide a consistent means for the distributed processing of grid simulations at a cloud scale.
The specific contributions of this paper are:
\begin{enumerate}
    \item Stochastic-aware distribution system planning application for PV and EV adoption by customers in a distribution grid. The application leverages a Markov model for DER adoption by customers, imposing both temporal and spatial uncertainty and a physics-based grid simulator (GridLAB-D) to conduct load flow analysis.
    \item Cloud-based architectural framework with modular components for scenario generation, grid simulation, and post-processing. The application leverages the scalability of cloud services to generate and analyze a full range of planning scenarios for risk-aware decision support.
\end{enumerate}

% \begin{figure}[t]
%     \centering
%     \includegraphics[width=0.5\textwidth]{images/framework.png}
%     \caption{Proposed Framework for DER Integration Study. \textbf{Just a placeholder}}
%     \label{fig:framework}
% \end{figure}

\section{Methodology}
\subsection{DER Adoption Model and Scenario Generation}
We begin by selecting a representative test feeder from the distribution system to initiate the stochastic analysis of DER adoption. Following a Markov-based adoption model, the population script is designed to distribute solar PV and EV across the feeder. This model utilizes predefined adoption probabilities to simulate consumers' gradual uptake of DER technologies over time (see Fig. 1).
% To model the probability of each customer adopting PV or EV technology in a given year, we employ a discrete-time Markov model. This approach leverages location-based adoption probabilities, providing insights into the likelihood of adoption across different years for each customer. 
% Our model is based on the following assumptions:
% \begin{enumerate}
%     \item Adoption probability \( p_i \) for each customer is location-specific and remains constant across years.
%     \item Once a customer adopts PV/EV, they remain in the adoption state permanently.
%     \item The decision to adopt is independent of the adoption status of other customers.
% \end{enumerate}

Let \( \mathcal{S} = \{0, 1\} \) represent the state space for each customer, where: \( 0 \) indicates that the customer has not adopted PV/EV technology and \( 1 \) indicates that the customer has adopted PV/EV technology.
Each customer \( i \) is associated with a location-based probability of adoption \( p_i \), which influences the transition probabilities between states.
We define the transition probability matrix \( P_i \) for customer \( i \) as follows:
\[
\small
P_i = \begin{bmatrix}
1 - p_i & p_i \\
0 & 1
\end{bmatrix}
\]
where: \( P_i(0 \rightarrow 1) = p_i \): the probability that customer \( i \) adopts PV/EV technology in a given year, given that they have not previously adopted it. \( P_i(1 \rightarrow 1) = 1 \): once a customer adopts PV/EV technology, they remain in the adopted state.

% In this work, we used the following probabilities for EV and PV adoption based on the demographic index ($d_{idx} = \{1, 2, 3, 4, 5\}$).
% \begin{equation}
% p_i(PV) = -(d_{idx} + 1)^{-\frac{1}{17}} + 1
% \end{equation}
% \begin{equation}
% p_i(EV) = -(d_{idx} + 1)^{-\frac{1}{13}} + 1
% \end{equation}

The state \( X_i(t) \) of customer \( i \) at year \( t \) evolves according to the Markov process. Given the initial state \( X_i(0) = 0 \), the probability that customer \( i \) adopts PV/EV by year \( t \) can be expressed as:
\vspace{-0.25 cm}
\[
P(X_i(t) = 1) = 1 - (1 - p_i)^t
\]
This formulation assumes a homogeneous probability \( p_i \) for each year based on location and customer-specific attributes.  Using our Markov model, we can generate long-term adoption scenarios to assess the evolution of DER adoption for the next \( n \) years. Each scenario represents a unique combination of location and DER capacity, allowing us to explore how adoption may unfold under different circumstances.
% To simulate DER adoption over \( n \) years, we define:
% \begin{itemize}
%     \item \( n \): the number of years for the simulation horizon (e.g., \( n = 30 \)).
%     \item \( m \): the number of trials for each location-based scenario, where each trial represents a unique adoption path due to the stochastic nature of customer decisions.
%     \item \( p_i \): the location-specific probability of adoption for customer \( i \), as described in the Markov model.
% \end{itemize}

For each customer \( i \) and each year \( t \), the state \( X_i(t) \) transitions based on the probability \( p_i \) of adopting PV/EV technology.
The process for generating scenarios is as follows:
\begin{enumerate}[noitemsep,topsep=0pt,leftmargin=*]
    \item \textbf{Initialize:} For each customer \( i \) without EV and PV at year 1, set the initial state \( X_i(0) = 0 \), indicating no adoption.
    \item \textbf{Yearly Simulation:} For each year \( t = 1, 2, \dots, n \):
        \begin{itemize}
            \item Update the state \( X_i(t) \) for each customer based on the transition probability \( p_i \).
            \item Track the cumulative DER capacity and location of adopted customers to capture the evolving DER landscape (see Fig. 2).
        \end{itemize}
    \item \textbf{Repeat:} Repeat the above process for \( m \) trials, where each trial represents a unique realization of the adoption process, capturing the variability in customer decisions.
\end{enumerate}
A total of \( n \times m \) scenarios are generated, representing different DER adoption patterns over time.

\subsection{Grid Simulation and Impact Analysis}
In the next step, we perform a power flow analysis on each generated scenario, capturing grid states that include bus voltages, line flows, and asset loading conditions. Once the power flow analysis is complete, we use post-processing scripts to analyze the Monte Carlo simulation results, aiming to identify and quantify EV/PV integration risks. For this study, we focus on service transformer overload as a key metric to assess grid impacts. 

From the $n\times m$ scenario, we evaluate the annual frequency of transformer overload. Two key metrics are chosen for visualizing these impacts: (i) the year when the first violation occurs and (ii) the frequency of violations within $m$ trials for each year $n$. Using this data, each service transformer can be represented as a series of concentric circles within the grid topology. Each circle corresponds to a different year, with the circle's size indicating the likelihood of impact in that year and its color showing the earliest an impact was detected.    
Finally, the user can generate insightful visualizations and plots illustrating the probability distributions of service transformer overloads, facilitating informed decision-making by distribution utilities.

\begin{figure}[t]
    \centering
    \includegraphics[width=0.425\textwidth]{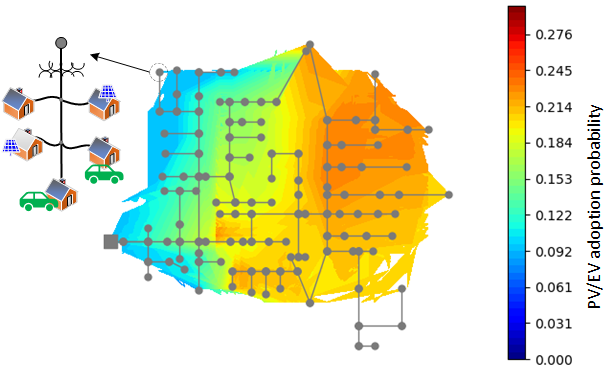}
    \vspace{-0.35 cm}
    \caption{IEEE 123-bus test feeder with adoption probability of customers. Each lumped load is disaggregated into a set of houses.}
    \label{fig:testcase}
    \vspace{-0.5 cm}
\end{figure}

\section{Proposed Architectural Framework}
\begin{figure}[t]
    \centering
    \includegraphics[width=0.495\textwidth]{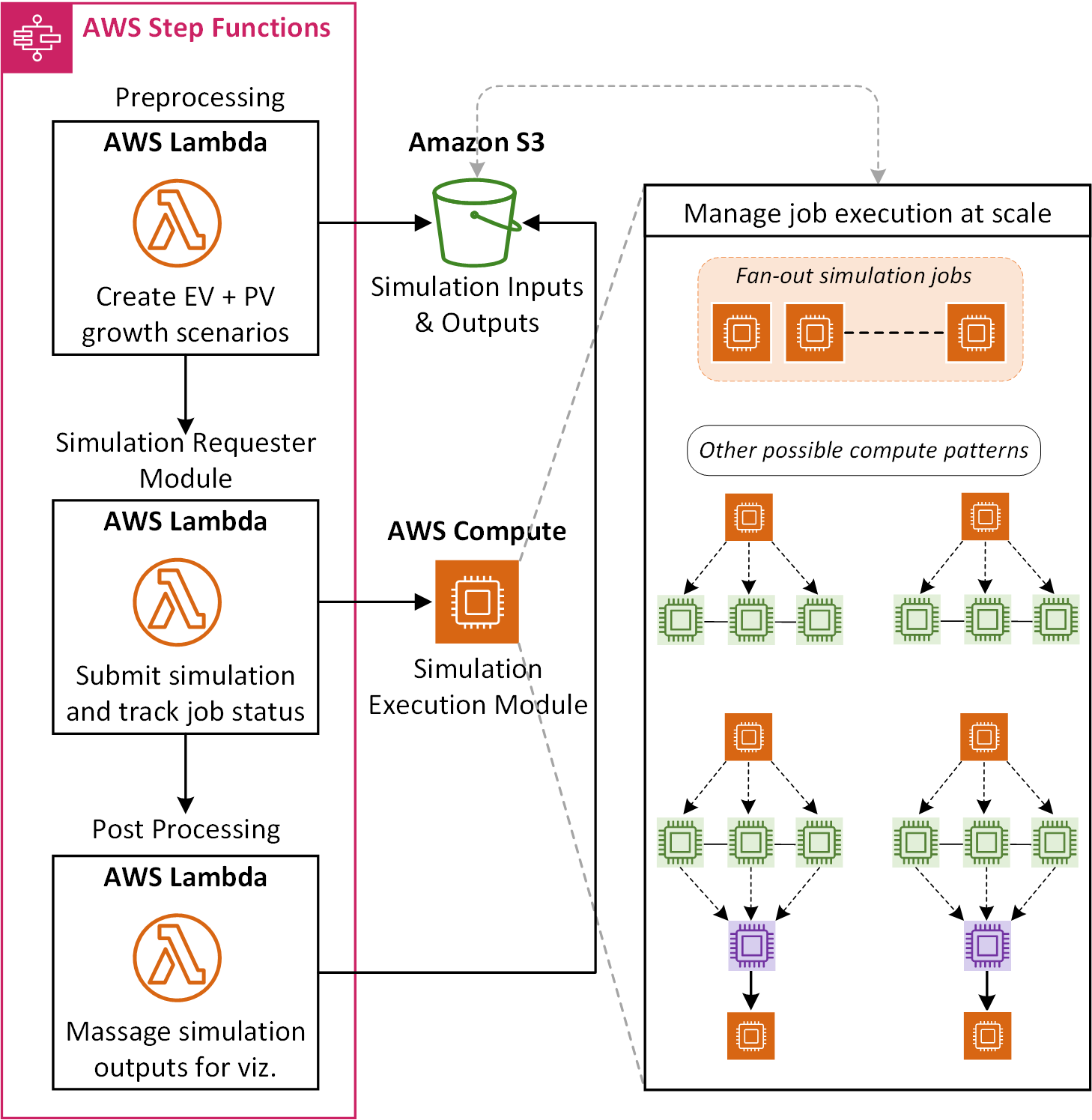}
    \vspace{-0.4 cm}
    \caption{Proposed architecture showing modular components orchestrated as a workflow (left) and various distributed processing compute patterns (right).}
    \vspace{-0.5 cm}
    \label{fig:aws}
\end{figure}

The proposed architectural framework deconstructs the methodology into three simple steps: 1) pre-processing to generate inputs required for the MC simulations, 2) execution of simulations at scale, and 3) post-processing results to identify the location, year of occurrence, and severity of anticipated grid violations. This decoupling of grid simulations from the rest of the use-case-specific logic allows for the future extensibility of this architectural framework. For example, new use cases can re-use simulation execution capabilities at a cloud scale, leveraging hundreds or thousands of compute nodes. Fig. \ref{fig:aws} shows a reference architecture for the three-step workflow using AWS Step Functions [https://aws.amazon.com/step-functions/], a visual workflow orchestration service that helps developers use AWS services to build distributed applications, automate processes, orchestrate microservices, and create data and machine learning (ML) pipelines.

% \begin{figure}[t]
%     \centering
%     \includegraphics[width=0.495\textwidth,trim={0.75cm 0.55cm 0.5cm 0.54cm},clip]{images/aws-archi.pdf}
%     \vspace{-0.4 cm}
%     \caption{Proposed architecture showing modular components orchestrated as a workflow (left) and various distributed processing compute patterns (right).}
%     \vspace{-0.5 cm}
%     \label{fig:aws}
% \end{figure}

\subsection{Pre-processing}
The three-step process begins with the generation of scenarios to be analyzed (preprocessing). AWS Lambda [https://aws.amazon.com/pm/lambda/], a serverless computing service that runs code in response to events, and automatically manages the compute resources, is used for scenario generation, as shown in Figure 1. AWS Step Functions integrates with AWS Lambda to invoke the scenario generation Lambda function, which executes Python code for generating simulation inputs. AWS Lambda allows PNNL researchers to focus on the business logic for scenario generation, such as the use of Markov-model for DER adoption, without thinking about servers or execution runtime. This scenario generation Lambda function generates data for $m$ trials, where each trial uses probabilities to model DER adoption and customer behavior for each hour over the next $n$ years. This leads to the creation of GridLab-D (.glm) and supporting files (weather and load profiles) per year per trial, representing the inputs for those $m$ trials worth of $n$ years hourly simulations. Here, each trial is independent of another and can be run concurrently with other trials. Within a trial, however, previous years’ DER adoption informs all the locations available for further DER adoption in successive years, and therefore, requires sequential execution. All files are stored in Amazon Simple Storage Service (S3) [https://aws.amazon.com/s3/].

\subsection{Simulation Execution}
The next step is the re-usable simulation execution part of the proposed architecture. This step is implemented in the form of a \textit{requester} module that submits simulation jobs to be executed and an \textit{execution} module that responds to these requests by executing them and allowing the \textit{requester} to check the status of previously submitted jobs. This allows the execution module to be re-used for other use cases. Another AWS Lambda function is used to implement the \textit{requester} module, submitting a yearly power flow simulation job for each year and each trial. For $m$ trials over $n$ years, this equals submitting $n\times m$ jobs, where each job needs to run a yearly simulation with 8,760 data points using GridLAB-D. Since all inputs required for each year and each trial are pre-computed in the previous step of the workflow, all simulation jobs can be submitted concurrently and executed independently. For queuing, execution, and management of these simulation jobs, AWS worked closely with PNNL to support running thousands of 8760 grid simulations. 
For this scaled execution, AWS contributed a cloud solution (diagram icon ``Simulation Execution Module''). This cloud solution supported the execution of different compute patterns, including a simple fan-out to a large number of distributed compute nodes, required for such stochastic analysis, as well as other, more complex, and multi-step compute patterns as depicted to the right of Fig. \ref{fig:aws}. The re-usable \textit{execution} module offers the performance and scale required for stochastic analyses that can model various uncertainties described in the previous sections. 

\begin{figure}[t]
    \centering
    \includegraphics[width=0.5\textwidth]{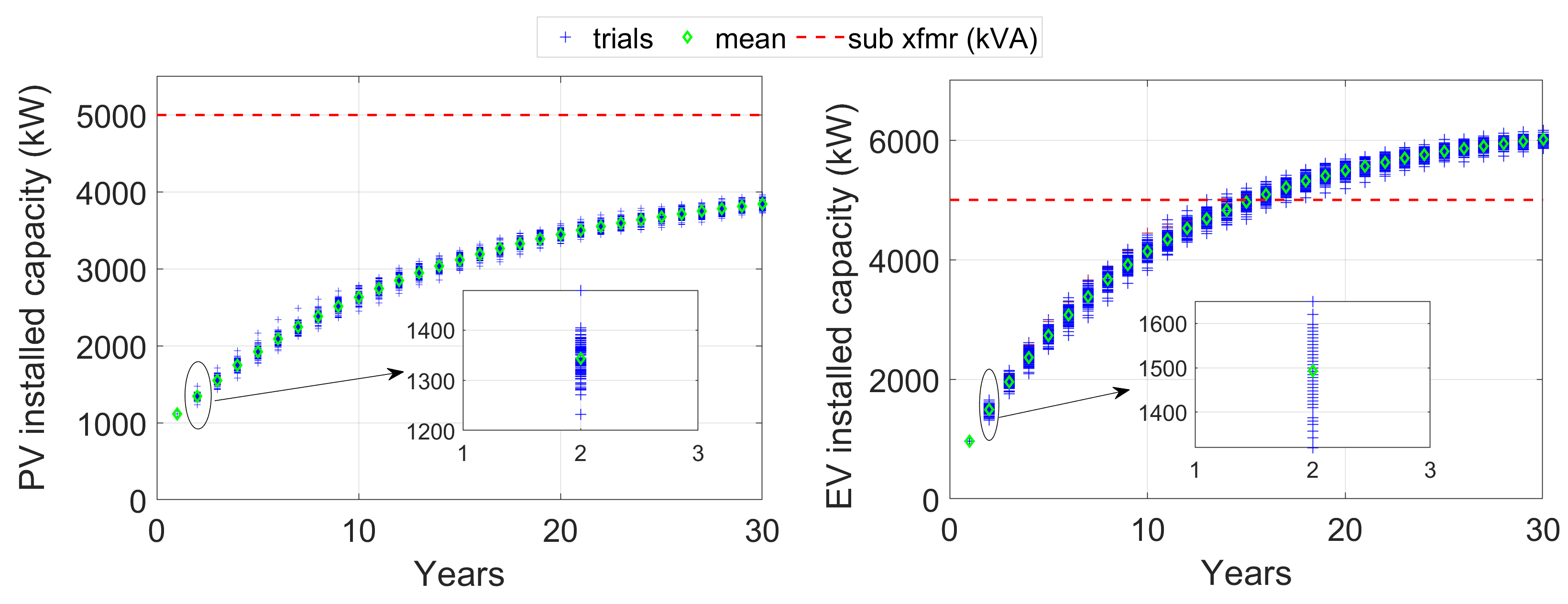}
    \vspace{-0.75 cm}
    \caption{Feeder PV/EV capacity for 500 $\times$ 30-year scenarios.}
    \vspace{-0.5 cm}
    \label{fig:scenarios}
\end{figure}

\begin{figure*}[t]
    \centering
    \includegraphics[width=0.9\textwidth]{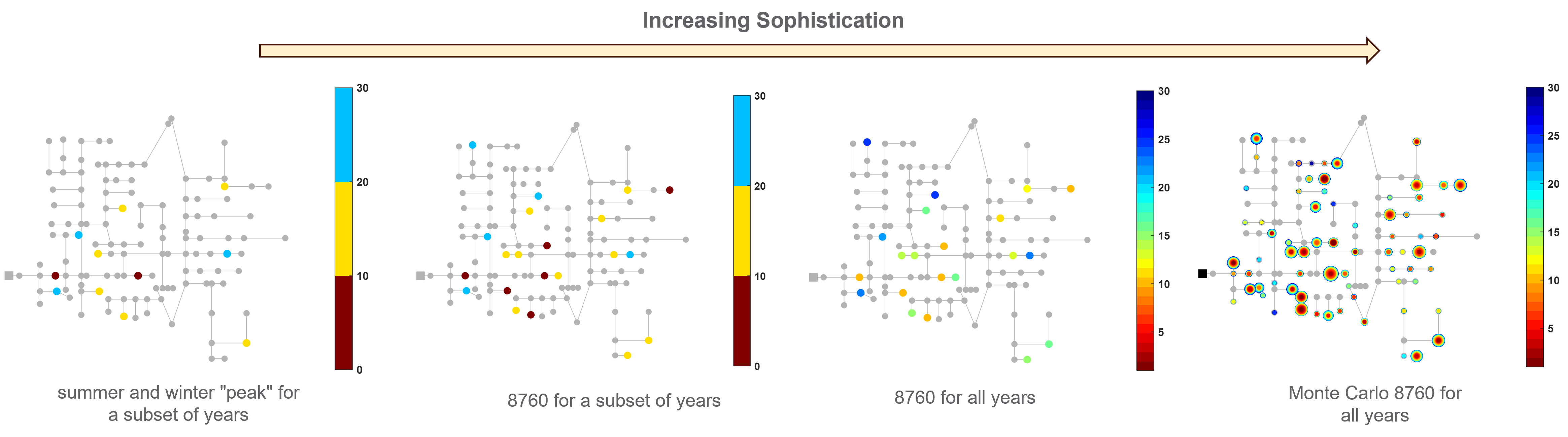}
    \vspace{-0.5 cm}
    \caption{Performance comparison with increasing sophistication in simulation.}
    \label{fig:comparison}
    \vspace{-0.5 cm}
\end{figure*}

\subsection{Post-processing}
The final step of the workflow is the postprocessing step, where results from 8,760 power flow analyses over $n$ years and $m$ trials are processed to identify voltage violations and asset overloads. This step filters and massages the simulation outputs so that a visualization can be easily developed to demonstrate the location of grid violations, the frequency of those violations over the numerous trials, and the timing and severity of those violations over the $n$ year, hourly, quasi-static time series simulations.  The proposed architectural framework supports running both Python or MATLAB, using an AWS Lambda function for Python code execution, or running MATLAB on AWS [https://aws.amazon.com/blogs/publicsector/matlab-parallel-cloud-computing-aws-researchers/]. 

\section{Demonstration}
% \subsection{Test Feeder}
For our study, we selected the IEEE 123 bus test feeder from the GridAPPS-D database as the basis for our demonstration. We modified the test feeder to investigate DER adoption models---enhancing its ability to accommodate customer-owned PVs and EVs by disaggregating lumped loads into a secondary model that reflects load distribution across the network (see Fig. \ref{fig:testcase}). This approach establishes customer points within the feeder for EV/PV integration, enabling a more granular analysis of adoption patterns and their impacts on distribution operations, such as transformer overloading.

\begin{figure}[t]
    \centering
    \includegraphics[width=0.495\textwidth]{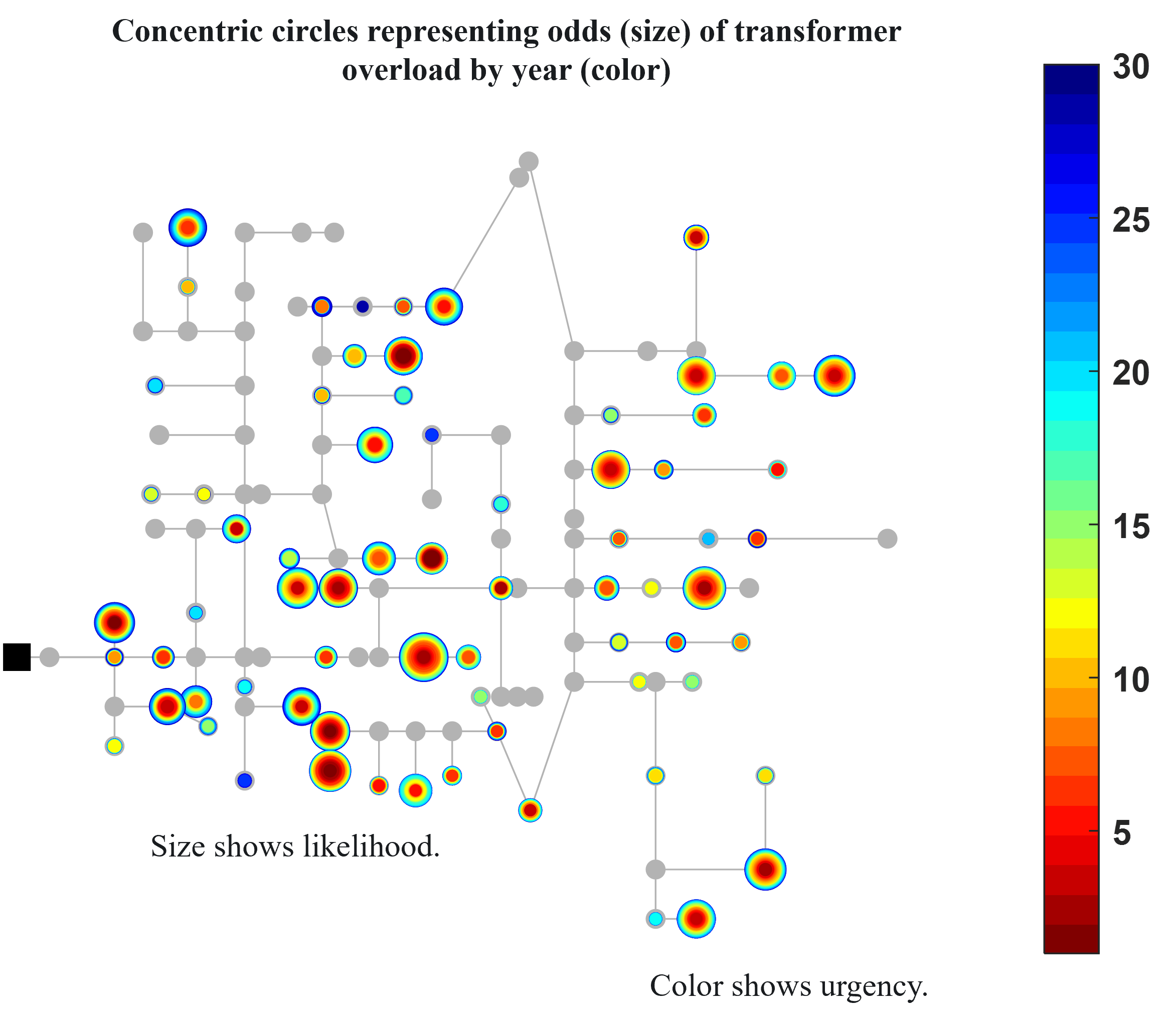}
    \vspace{-0.75 cm}
    \caption{Impact visualization in service transformers due to potential EV/PV growth in 30 years.}
    \label{fig:visual}
    \vspace{-0.45 cm}
\end{figure}

\subsection{Simulation Setup}
For the demonstration, 500 trials are conducted for a 30-year growth model of customer-owned EV/PV (i.e., 15,000 yearly simulations with 8,760 timestamps).  Fig. \ref{fig:scenarios} shows the growth model for PV and EV for the next 30 years.  Once the models and data representing this EV and PV growth are generated, and compute resources are provisioned for executing grid simulation jobs, the \textit{requester} module submitted jobs containing a list of power flow requests. Each request included pre-processed files, including a GridLAB-D model, weather data, and load profiles for 8,760 timestamps in a zip format. The \textit{requester} module also specified the Amazon S3 location where the job output should be stored for post-processing. The executor module verified the validity of the pre-processed files and then started the GridLAB-D simulation for each request. The 15,000 power flow jobs are created and submitted concurrently. Once all the jobs were submitted, the \textit{requester} module periodically checked for completion. The \textit{execution} module processed 1,000 jobs in parallel. An even higher level of parallel execution can be used if needed, either to process the jobs even faster or to execute more analyses in the same amount of time. 

\subsection{Results}

Fig. \ref{fig:comparison} shows a performance comparison with traditional approaches where DER adoption is studied under particular scenarios. For example, some conventional approaches use worst-case scenarios such as summer peak and winter peak \cite{cicilio2020transmission}, and the business-as-usual distribution system capacity planning horizons are typically five years \cite{keen2022distribution}. 
With such approaches, outlier results cannot be identified; potential grid impact conditions might be missing and lead to excessively conservative conclusions that constrict DER interconnection. 
As we move from left to right in Fig. \ref{fig:comparison}, it becomes apparent that expanded scenario analysis, drawing from rich datasets, captures a more comprehensive grid response, bolstering confidence in the resulting conclusions.
The increase in sophistication (i.e., scaling DER growth with probabilistic studies) allows the operator to understand better where/when the problems will occur.
\vspace{-0.2 cm}
\begin{figure}[h]
    \centering
    \includegraphics[width=0.45\textwidth]{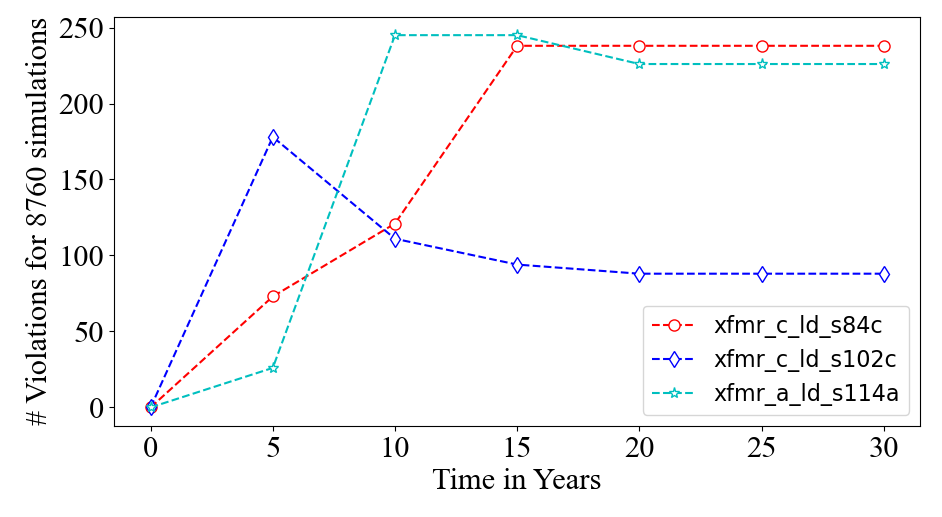}
    \vspace{-0.4 cm}
    \caption{Number of violations in a yearly simulation for selected transformers}
    \vspace{-0.15 cm}
    \label{fig:violations}
\end{figure}

Fig. \ref{fig:visual} shows a visualization of service transformer overload with concentric circles representing the impact (odds [size of the circle] of transformer overload by year [color of the circle]). The size of the circle shows the likelihood that problems start to occur, and the circle's color shows urgency (i.e., transformer overloads in the early years of the DER adoption). For instance, a large red circle represents that the problem in a transformer will occur in the early years, and the problem occurred during most of the trials.

We also studied a few trials to analyze the impact of the co-adoption of EVs and PVs. The impacts of these two technologies may not be additive, and EV charging behaviors may be subject to change after solar panels are added. Consumers may adapt EV charging to the variable solar energy generation
and shift EV charging to the hours when solar panels generate
electricity \cite{liang2022impacts}. 
While we do not specifically model the change in customer behavior after adopting such technologies, we randomly selected trials to observe this. Fig. \ref{fig:violations} shows a plot where the number of violations of service transformer is reported every five years for the next 30 years. In certain trials, it is observed that the frequency of overloads in transformers decreases after 10-15 years despite the increasing adoption of EVs. This phenomenon can be attributed to the expected shift in electricity consumption patterns among EV users who also adopt solar panels, leading to a change in their energy usage behavior.
Such findings could help with load management strategies (i.e., smart charge management in EVs) and infrastructure upgrades.

\begin{figure}[t]
    \centering
    \includegraphics[width=0.42\textwidth]{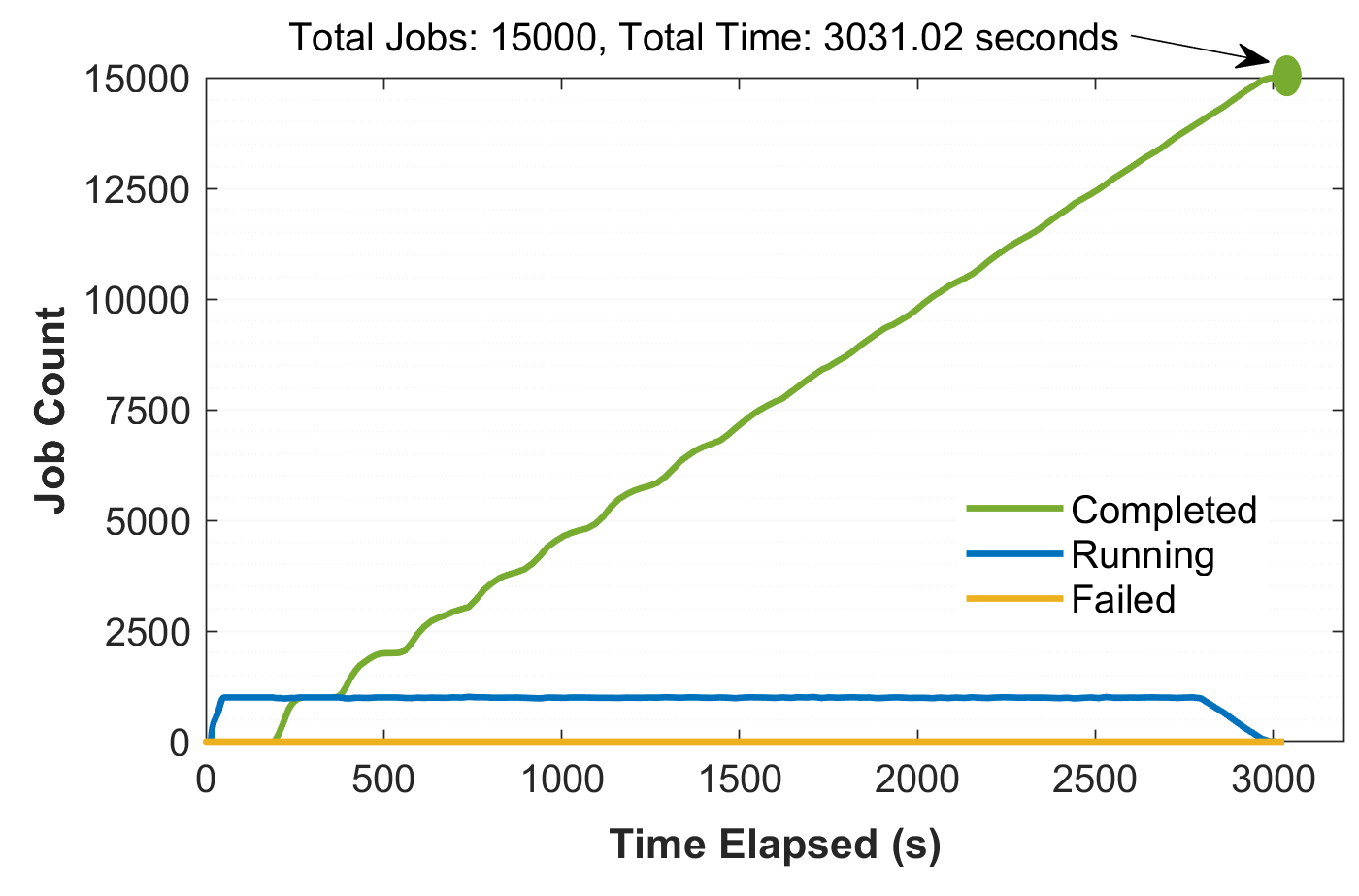}
    \vspace{-0.35 cm}
    \caption{\textcolor{black}{Monitoring execution performance for 15,000 jobs, each representing an 8760 powerflow simulation.}}
    \label{fig:execution}
    \vspace{-0.25 cm}
\end{figure}

To showcase the scalability, we monitored the execution time of power flow jobs. The total number of monitored, completed, and failed jobs are tracked each time. Fig. \ref{fig:execution} shows the job completion rate for the execution of 15,000 simulations---each covering all 8,760 hours representing a yearly simulation---demonstrates that a total of 131.4 million power flow calculations (8,760 × 15,000) were completed in under an hour. 
% \textcolor{blue}{In contrast, a single yearly simulation on a local PC [32 GB RAM; 13th Gen Intel(R) Core(TM) i9] took approximately 30 minutes. Running 15,000 such simulations within a reasonable timeframe would require job distribution across multiple computing resources, consequently increasing in-house infrastructure costs.}
\textcolor{black}{In contrast, an engineering workstation [32 GB RAM; 13th Gen Intel(R) Core(TM) i9] completed one job in around 32 minutes. A performance summary is shown in Table~\ref{cloud}.}
% In addition to scalability, the deconstructed architecture allows for easy re-use of its components for other use cases. Reusable and easy access to high-scale, low-cost technologies encourage experimentation and innovation. 
In addition to scalability, the deconstructed architecture enables easy reuse of its components for other use cases, providing accessible, high-scale, low-cost technologies that encourage experimentation and innovation. For instance, the project team is adopting these architectural patterns and AWS technologies for new use cases in distribution operations.  

% \begin{figure}[t]
%     \centering
%     \includegraphics[width=0.42\textwidth]{images/metrics_aws.eps}
%     \vspace{-0.35 cm}
%     \caption{\textcolor{black}{Monitoring execution performance for 15,000 jobs, each representing an 8760 powerflow simulation.}}
%     \label{fig:execution}
%     \vspace{-0.25 cm}
% \end{figure}

\begin{table}[t]
    \centering
    \caption{\textcolor{black}{Performance Comparison for 15,000 Jobs: Cloud Solution vs. Engineering Workstation}}
    \begin{tabular}{|l|c|c|}
    \hline
        \multirow{1}{*}{Computing resource} & Avg. time per job& Total execution time \\
        \hline
        Cloud solution & 0.202 seconds & 3031.02 seconds\\
        Engineering workstation & 32.12 minutes  & 334.5 days  \\
        \hline
        % \multicolumn{3}{|c|}{*[32 GB RAM; 13th Gen Intel(R) Core(TM) i9]} \\
        % \hline
    \end{tabular}
    \vspace{-0.5 cm}
    \label{cloud}
\end{table}

\section{Conclusions}
This paper presented a stochastic-aware distribution planning application that addresses uncertainties arising from the growing adoption of EVs and PVs. By incorporating temporal and spatial variability using a Markov model and leveraging a cloud-based architecture, our approach facilitates efficient large-scale scenario generation and grid impact simulations. Moreover, our analyses of large-scale datasets provided key insights into the spatial and temporal impacts of DERs, underscoring the limitations of planning based on average conditions alone and revealing that worst-case scenario planning, while robust, may introduce high costs. Furthermore, the use of managed cloud services further highlights the benefits of scalable, high-performance computing without the need for costly, in-house supercomputing infrastructure, streamlining implementation and reducing internal development efforts.

% https://www.nrel.gov/docs/fy23osti/83892.pdf
% https://www.nrel.gov/docs/fy23osti/84543.pdf
% https://www.nrel.gov/docs/fy22osti/81811.pdf
% https://www.nrel.gov/docs/fy17osti/68869.pdf
% https://www.epri.com/research/products/000000003002011009
% https://www.sciencedirect.com/science/article/pii/S0140988322003231
% https://www.nature.com/articles/s41560-017-0074-z

\bibliographystyle{IEEEtran}
\balance
\bibliography{references.bib}
% \bibliography{main.bbl}
\balance

\end{document}